\documentclass[a4paper,11pt]{article}
\usepackage{pos}

\usepackage{slashed}

\newcommand{\Dcov}{D}
\newcommand{\Dslash}{\slashed{\Dcov}}
\newcommand{\eps}{\epsilon}
\newcommand{\Pf}{\operatorname{Pf}}

\newcommand{\sigmabar}{\bar{\sigma}}

\newcommand{\ep}{\epsilon}
\newcommand{\ept}{\tilde{\epsilon}}

\newcommand{\psib}{\bar{\psi}}

\newcommand{\gammab}{\bar{\gamma}}

\newcommand{\lambdab}{\bar{\lambda}}

\newcommand{\im}{\mathrm{i}}
\newcommand{\esi}{\left(\epsilon\cdot\sigma\right)}
\newcommand{\esib}{\left(\epsilon\cdot\bar{\sigma}\right)}

\hypersetup{pdftitle=Charged massive spin 2 and 3/2 propagation in a constant electromagnetic background}
\hypersetup{pdfauthor={Karim Benakli, Wenqi Ke, Bruno Le Floch}}
\title{Charged massive spin 2 and 3/2 propagation in a constant electromagnetic background}

\author*[a]{Karim Benakli}
\author[b]{Wenqi Ke}
\author[a]{Bruno Le Floch}

\affiliation[a]{Sorbonne Universit\'e, CNRS, Laboratoire de Physique Th\'eorique et Hautes Energies, LPTHE, \\F-75005 Paris, France}

\affiliation[b]{William I. Fine Theoretical Physics Institute, School of
 Physics and Astronomy, University of Minnesota,\\
Minneapolis, MN 55455, USA}

\emailAdd{kbenakli@lpthe.jussieu.fr}
\emailAdd{wke@umn.edu}
\emailAdd{blefloch@lpthe.jussieu.fr}

\abstract{We present two methods for deriving the equations of motion for charged massive spin-3/2 particles. The first approach involves utilizing the Euler-Lagrange equations derived from a Lagrangian that describes the propagation of the first massive excitation of open superstrings. The second method entails enforcing the conditions that the trace vanishes and that the covariant derivative of the equations vanish. We very briefly comment about other spins.}

\FullConference{Corfu Summer Institute 2023 "School and Workshops on Elementary Particle Physics and Gravity" (CORFU2023)\\
 April 23 to May 6, and August 27 to October 1, 2023\\
Corfu, Greece\\}

\begin{document}
\maketitle

\section{Introduction}

The elementary particles that are currently known are believed to be composed of fundamental states with different spins: spin~0 for the Higgs (although it is still under investigation whether it is composite or not), spin~1/2 for matter fields such as quarks and leptons, spin~1 for the Standard Model gauge bosons, and spin 2 for gravitons. A spin~3/2 state is missing from this enumeration. In fact, we do not know if fundamental particles with spin~3/2 actually exist in nature. A hypothetical candidate spin 3/2 particle is  the gravitino, the supersymmetric partner of the graviton. But as most of the energy in our Universe is of an unknown form, it remains possible that there could be massive, and maybe charged, higher-spin fundamental particles. While such fundamental states have not been observed in Nature, composite states with spin higher than one, massive and charged, do exist. Hadronic resonances are an example of such states. In the case of spin~3/2, we have two often-cited examples: the $\Delta$ resonance, which can be seen as the result of flipping one of the quarks' spin inside a proton, and the $\Omega^-$, which played a historical role in the discovery of the quark model. The $\Delta$ resonance is extremely short-lived, but the $\Omega^-$ appears to propagate a few centimeters in a bubble chamber (thus a charged spin 3/2 state in a constant electromagnetic background). One might try to consider it as a ``localized'' particle and describe its propagation via an effective field theory.

The study of high-spin particles in field theory has posed a formidable and persistent challenge in physics.  Particularly in the case of massive particles, the task of propagating states of spin greater than~1 in an electromagnetic background turned out to be a daunting one. As early as 1936, Dirac emphasized the need for formulating equations of motion for these states~\cite{Dirac:1936tg}. Fierz and Pauli took up the challenge and in their 1939 work~\cite{Fierz:1939ix} yielded the now-famous Fierz-Pauli Lagrangian for a neutral massive spin 2 particle. Still many issues remain open problems. The challenges associated with formulating field theories for high spin particles have led to wonder whether these difficulties are a sign that such particles do not exist in nature. However, efforts to construct field theories for these hypothetical particles might allow us to gain a deeper understanding of their possible properties,  to comprehend why such states may or may not exist in our universe,  and to gain a greater appreciation for the underlying principles that govern the building blocks of our universe.

Several decades after Dirac, Fierz and Pauli, it became clear that the difficulty in building a field theory for higher spin particles was more profound than initially thought. Johnson and Sudarshan~\cite{Johnson:1960vt}, then Velo and Zwanziger~\cite{Velo:1969bt,Velo:1969txo,Velo:1972rt}, made significant contributions in this regard. In particular, when trying to canonically quantize minimally-coupled spin-3/2 fields, Johnson and Sudarshan discovered that the equal-time commutators were incompatible with the relativistic covariance of the theory. Velo and Zwanziger went on to demonstrate that the minimally coupled Lagrangians for spin 3/2 and spin 2 fields exhibited pathological behavior at the classical level. Interestingly, both problems emerged at a specific value of the electromagnetic field strength, suggesting a common underlying issue. It was later realized that the set of secondary constraints became degenerate, which is a sign of the loss of invertibility. This loss of invertibility means that the constraints no longer determine all the components of the fields, leading to a breakdown of causality and hyperbolicity. Another way the problem appears in the construction of Lagrangians that lead to systems of Fierz-Pauli equations for high spin particles, is through the additional fields of lower spin $s-1, s-2, \dots$. In the free case, these fields are required non-propagating auxiliary fields. However, in known tentative Lagrangians, the different components of the higher spin fields are mixed in a non-trivial way with these fields, resulting in the mixing of physical and auxiliary fields, leading to non-causal propagation. Unfortunately, the only known {\it four-dimensional Lagrangian} that describes an isolated charged massive spin 2 state, the Federbush Lagrangian~\cite{federbush_1961}, suffers from causality loss as it has superluminal propagating modes. Currently, there is no satisfactory fully explicit Lagrangian that describes just a massive charged spin-3/2 particle in a way that is theoretically consistent, though an implicit form was proposed in~\cite{Porrati:2009bs}. There, an ansatz for the Lagrangian was written where the coefficients of the different terms can be obtained recursively order by order in the electromagnetic field strength.

String theory exhibits states with arbitrarily high spin in the Regge trajectories. Therefore, after the question of describing the propagation of strings in an electromagnetic field was solved by~\cite{Abouelsaood:1986gd,Burgess:1986dw,Nesterenko:1989pz}, Argyres and Nappi employed String Field Theory to investigate the case of the first massive level of the open bosonic string in a constant electromagnetic background \cite{Argyres:1989cu,Argyres:1989qr}, and succeeded in deriving a Lagrangian for the massive charged spin 2 field. Though undoubtedly a great success, this Lagrangian unfortunately suffers from pathologies in any dimension other than $d=26$. Subsequently, Porrati and Rahman investigated its reduction to four dimensions \cite{Porrati:2011uu}, and demonstrated that it yields a spin 2 field coupled to a scalar. Despite many efforts, the problem of constructing a Lagrangian that only contains fields of higher spins remains unresolved to this day. However,  progress  was made for the original problem posed by Dirac, Fierz and Pauli of simply writing down the equations of motion  \cite{Benakli:2023aes,Benakli:2022edf,Benakli:2022ofz,Benakli:2021jxs} which we will be concerned with here.

\section{Deriving spin-3/2 equations of motion  from a Lagrangian}

In light of the challenges encountered in formulating a self-consistent set of equations of motion and constraints for charged massive higher spins, Fierz and Pauli \cite{Fierz:1939ix} proposed obtaining them from a Lagrangian framework. String theory has emerged as particularly valuable in this regard, thanks to the presence of a higher spin tower in its spectrum. In \cite{Benakli:2021jxs} an effective superspace action for the first mass level of a charged superstring, incorporating a massive spin-3/2 and a massive spin-2 particle within the same supermultiplet, was  obtained. This action was subsequently expanded and simplified into its component form \cite{Benakli:2023aes,Benakli:2022edf,Benakli:2022ofz}, which retains only physical degrees of freedom. The resulting Lagrangian includes a charged massive spin-3/2, described\footnote{We use indices $m,n,p$ as spacetime indices while $i,j$ are spatial indices.} by the two-component spinor fields ${\lambda}^m, {\chi}_{m}$,  coupled to the two-component spinors $\psi, \gamma$ of a spin 1/2  Dirac fermion\footnote{More precisely, the string theory first excited level contains two copies of these fermions, and their couplings to the electromagnetic background differ by minus signs \cite{Benakli:2023aes,Benakli:2022edf,Benakli:2022ofz}},
\begin{equation}
    \begin{aligned}
    \mathcal{L}_F=&-\frac{\mathrm{i}}{2}\left[( \lambda^m \sigma^n\mathfrak{D}_n\bar{\boldsymbol{\lambda}}_{m})  
    +(\bar{\boldsymbol\chi}_{m} \bar{\sigma}^n{\sigma}^k\bar{{\sigma}}^m\mathfrak{D}_k\boldsymbol\chi_{n})\right]-\left[ ({\boldsymbol{\lambda}}^m{\boldsymbol{\chi}}_{m})+\text{h.c.}\right] 
\\&+30\mathrm{i}(\psi\sigma^m \mathfrak{D}_m\psib )+2\mathrm{i}( \gamma\sigma^m\mathfrak{D}_m\gammab)
\\&+ \left[-3\mathrm{i}({\boldsymbol{\chi}}^m \sigma_n\sigmabar_m\mathfrak{D}^n \psi)-3( \boldsymbol{\lambda}^m\sigma_m\psib)-2\mathrm{i}(\bar{\boldsymbol{\chi}}^m \sigmabar_m\gamma)-(\boldsymbol{\lambda}^m\mathfrak{D}_m \gamma)+\text{h.c.} \right]\\&+\bigl[18\mathrm{i}( \psi\gamma)+\text{h.c.}\bigr]-\frac{1}{2}\bigl[\bar{\boldsymbol{\chi}}^m (\epsilon\cdot\bar{\sigma})\bar{\sigma}_m\gamma+\text{h.c.}\bigr] ,
\end{aligned}
\label{physLF}
\end{equation}
with notation given momentarily.
The Lagrangian describes first excitation level modes of open superstrings that carry total charges $Q = q_0 + q_\pi$, and involves dressed covariant derivatives~$\mathfrak{D}_m$ with commutator~\cite{Abouelsaood:1986gd,Argyres:1989cu}
\begin{equation}
  \left[\mathfrak{D}_m , \mathfrak{D}_n \right] =\im \epsilon_{mn} ,
  \label{mathfrakD_n}
\end{equation}
in terms of a dressed field strength $\epsilon_{mn}$ that encodes the field strength~$F_{mn}$.
Notably, our analysis continues to hold even if we take the limit $\epsilon_{mn} \rightarrow Q F_{mn}$ and $\mathfrak{D}_m \rightarrow D_m$.
The Lagrangian features the notation $(\eps\cdot \sigma)\equiv \eps^{mn}\sigma_{mn}$, $(\eps\cdot \sigmabar)\equiv \eps^{mn}\sigmabar_{mn}$ and rescaled spinors with bold symbols
\begin{equation}\begin{aligned}
&\bar{    \boldsymbol{\lambda}}_{m}\equiv \left( \eta_{mn}-\mathrm{i}\epsilon_{mn}\right)\bar{\lambda}^n ,\quad      \boldsymbol{\chi}_{m}\equiv \left( \eta_{mn}-\mathrm{i}\epsilon_{mn}\right)\chi^n .
\end{aligned}
\end{equation}

This Lagrangian leads to the  equations of motion
\begin{equation}
\begin{aligned}
  \im\bigl(\sigma^n\mathfrak{D}_n\boldsymbol{\bar{\lambda}}_{m}\bigr)_\alpha
  & = -\bigl(\eta_{mn}-\im\ep_{mn}\bigr)\Bigl(\sqrt{2}\boldsymbol{\chi}_{1}^n+(\im/2)\sigma^n\psib+\sqrt{2}\,\mathfrak{D}^n\gamma\Bigr)_\alpha ,
  \\
  \im\bigl(\sigmabar^n\sigma^k\sigmabar_m\mathfrak{D}_k\boldsymbol{\chi}_{n} \bigr)^{\dot{\alpha}}
  & = -2\sqrt{2}\boldsymbol{\lambdab}_{m}^{\dot{\alpha}}+3\sqrt{2}\left(\sigmabar_{mn}\mathfrak{D}^n\psib\right)^{\dot{\alpha}}-\mathfrak{D}_m\psib^{\dot{\alpha}}/\sqrt{2}
  \\&\quad -4\mathrm{i}\left(\sigmabar_m\gamma\right)^{\dot{\alpha}} -\left[\left(\epsilon\cdot\sigmabar \right)\sigmabar_m\gamma\right]^{\dot{\alpha}} ,
  \\
  \im\left(\sigma^m\mathfrak{D}_m\psib\right)_\alpha
  & = -6\sqrt{2}\left(\sigma_{mn}\mathfrak{D}^m\boldsymbol{\chi}^n\right)_\alpha+\sqrt{2}\,\mathfrak{D}^m\boldsymbol{\chi}_{1m\alpha}+2\im\left(\sigma^m\boldsymbol{\lambdab}_{m} \right)_\alpha+2\sqrt{2}\,\gamma_{\alpha} ,
  \\
  \im\left(\sigmabar^m\mathfrak{D}_m\gamma \right)^{\dot{\alpha}}
  & = -\im\left(\sigmabar^m\boldsymbol{\chi}_{m}\right)^{\dot{\alpha}}-\frac{1}{\sqrt{2}}\mathfrak{D}^m\boldsymbol{\lambdab}_{m}^{\dot{\alpha}} -\frac{1}{4}\left[\sigmabar^m\left(\epsilon\cdot\sigma \right)\boldsymbol{\chi}_{m} \right]^{\dot{\alpha}} -\frac{1}{2\sqrt{2}}\psib^{\dot{\alpha}} .
\end{aligned}
\end{equation}

Upon algebraic manipulations and performing the field redefinitions
\begin{equation} 
{\begin{aligned}
       \boldsymbol\lambdab^{\prime}_{m}&\equiv\boldsymbol\lambdab_{m}+\frac{\im}{2\sqrt{2}}\bigl[1-\im\esib\bigr]\sigmabar_m\gamma-\frac{1}{2} \bigl[\eta_{mn}-\im\left(\epsilon_{mn}+\im\ept_{mn}\right)\bigr]\mathfrak{D}^n\psib
, \\\boldsymbol{\chi}_{m}^\prime&\equiv \boldsymbol{\chi}_{m}+\frac{1}{2\sqrt{2}}\esi\sigma_m\psib,
        \end{aligned}}
\label{newspin32-indice1}
\end{equation}
we obtain the equation of motion and constraints
\begin{equation}
\begin{aligned}   
\bigl(\im \slashed{\mathfrak{D}}+\sqrt{2}\bigr)\boldsymbol{\Psi}_{m} & = \sqrt{2}\im \epsilon_{mn}\boldsymbol{\Psi}_{L}^n ,
\\
  \mathfrak{D}^m \boldsymbol{\Psi}_{m} & = \frac{1}{2\sqrt{2}}\left(\epsilon^{mn}
+\im\ept^{mn}\right)\gamma_n \boldsymbol{\Psi}_{m} ,
\\
  \gamma^m\boldsymbol{\Psi}_{m} & =0 , \end{aligned}  
\label{spin32 result}
\end{equation}
where we have arranged the two-component spinors into one Dirac spinor: 
\begin{equation}
  \boldsymbol{\Psi}_{m}\equiv \begin{pmatrix}\boldsymbol{\chi}^\prime_{m\alpha} \\
\bar{\boldsymbol{\lambda}}_{m}^{\prime\dot{\alpha}}
\end{pmatrix} .
\end{equation}
The spin 1/2 fields satisfy the familiar Dirac equation and need not to be displayed here.

\section{Deriving spin-3/2 field equations of motion without a Lagrangian}

We now describe a way to bypass arduous superstring calculations by directly determining a consistent set of evolution and constraint equations.  This leads to a class of equations that includes~\eqref{spin32 result} as a possibility.
We seek a Dirac equation of motion for the spin-3/2 field of the form
\begin{equation}
  \Dslash\Psi_m = \im M_{mn} \Psi^n
\label{equ-generic-form}
\end{equation}
where $\Dslash=\gamma^m\Dcov_m$ involves the covariant derivative operator~$\Dcov_m$ and the Dirac matrices~$\gamma_n$.

\begin{itemize}
\item We consider flat Minkowski space with metric $\eta\sim(-1,1,1,1)$, in particular $\Dcov_m\gamma_n=0$ and $\Dcov_m\eta_{np}=0$.  Our sign conventions are that $\{\gamma_m,\gamma_n\}=-2\eta_{mn}$, the Levi--Civita tensor has $\varepsilon^{0123}=1$, and $\gamma^5=\gamma^0\gamma^1\gamma^2\gamma^3$ obeys $\gamma^5\gamma^5=-1$ and $\gamma^{mnpq}=\varepsilon^{mnpq}\gamma^5$.  The left/right handed projectors are $P_L=(1+\im \gamma^5)/2$ and $P_R=(1-\im\gamma^5)/2$.  The dual field strength is $\ept^{pq} \equiv \frac{1}{2} \varepsilon^{mnpq} \eps_{mn}$.

\item We assume that $M_{mn}$ is a combination of gamma matrices and of the tensor~$\eps$ encoding the background field strength (and $\eta$, $\varepsilon$).

\item We restrict our analysis to the case of constant background $\eps$. This implies $\Dcov\eps=0$, therefore $\Dcov M=0$.

\item We impose $\gamma^m \Psi_m = 0$. 
This important projection ensures that the field $\Psi_m$ lies in the appropriate representation of the Lorentz group.
Labelling irreducible representations of $\mathfrak{so}(1,3)_{\mathbb{C}}=\mathfrak{sl}(2,\mathbb{C})\otimes\mathfrak{sl}(2,\mathbb{C})$ by a pair of spins, the unconstrained field~$\Psi_m$ transforms in
\begin{equation}
  (1/2,1/2)\otimes((1/2,0)\oplus(0,1/2)) = (1,1/2)\oplus(0,1/2)\oplus(1/2,1)\oplus(1/2,0) .
\end{equation}
The trace constraint $\gamma^m\Psi_m=0$ projects out the second and fourth summands, thus correctly eliminates the spin~1/2 part of the field.
This unmodified trace constraint is consistent with what superstring theory gave us in the previous section.
\end{itemize}

The (primary) trace constraint must be preserved by the time evolution given by the Dirac equation: this leads to the (secondary) divergence constraint\footnote{Strictly speaking this equation includes a time derivative, which should be cancelled by the evolution equation of~$\Psi_0$ to really obtain a constraint on the initial data.}
\begin{equation}
  0 = \Dslash(\gamma^m\Psi_m)
  = \{\gamma^n,\gamma^m\}\Dcov_n\Psi_m - \gamma^m\Dslash\Psi_m
  = -2 \Bigl( \Dcov^m\Psi_m + \frac{\im}{2} \gamma^m M_{mn} \Psi^n\Bigr) .
\end{equation}
This constraint itself must be preserved using only the Dirac equation and the trace and divergence constraints, as there are no further constraints in the vanishing background limit.
We find
\begin{equation}\label{D-on-equ-generic-form}
  \begin{aligned}
    0&= \Dslash\bigl(\Dcov^m \Psi_m + \frac{\im}{2} \gamma^m M_{mn} \Psi^n\bigr)
    \\
    & = \gamma_l [\Dcov^l,\Dcov^m] \Psi_m
    + \Dcov^m \Dslash \Psi_m
    + \frac{\im}{2} \gamma^l \gamma^m M_{mn} \Dcov_l \Psi^n
    \\
    & = \im \eps^{lm} \gamma_l \Psi_m
    + \im \Bigl( M_{mn} + \frac{1}{2} \gamma_m \gamma^p M_{pn} \Bigr) \Dcov^m \Psi^n .
  \end{aligned}
\end{equation}
The term involving $\Dcov^m \Psi^n$ must simplify using the evolution and constraint equations, hence the matrix multiplying it must include factors of $\gamma_m$ or $\gamma_n$ or $\eta_{mn}=-\{\gamma_m,\gamma_n\}/2$.
Altogether we can parametrize the possible matrices as (with a convenient normalization)
\begin{equation}\label{Mmnplusbla}
  M_{mn} + \frac{1}{2} \gamma_m \gamma^p M_{pn} = X_n\gamma_m+Y_m\gamma_n
\end{equation}
for two collections of matrices $X_n,Y_m$.  Contracting with $-\gamma^m$ yields
$\gamma^p M_{pn} = - \gamma^p (X_n\gamma_p+Y_p\gamma_n)$ hence
\begin{equation}\label{Mmn-X}
  M_{mn} = \frac{1}{2} \gamma_m \gamma^p (X_n\gamma_p+Y_p\gamma_n) + X_n\gamma_m+Y_m\gamma_n
  = - \frac{1}{2} \gamma^p \gamma_m X_n\gamma_p + \Bigl(Y_m + \frac{1}{2} \gamma_m \gamma^p Y_p\Bigr) \gamma_n .
\end{equation}
The last term, of the form $(\dots)\gamma_n$ plays no role because it drops out from the Dirac equation once taking into account the trace equation $\gamma_n\Psi^n=0$.  Thus we take $Y=0$ without loss of generality.
We then continue the calculation \eqref{D-on-equ-generic-form} using the explicit form~\eqref{Mmnplusbla} of~$M_{mn}$,
\begin{equation}
  0 = \im \eps^{lm} \gamma_l \Psi_m + \im X_n \Dslash \Psi^n
  = \bigl( \im \eps^{lp} \gamma_l - X_n M^{np} \bigr) \Psi_p
  = \bigl( \im \eps^{lp} \gamma_l + (1/2) X_n \gamma^q \gamma^n X^p \gamma_q \bigr) \Psi_p .
\end{equation}
This must not be a new constraint, hence it must be a multiple of the trace constraint.
We thus seek matrices $X_n$ and a new matrix~$Z$ such that
\begin{equation}\label{X-cond}
  X_n \gamma^q \gamma^n X^p \gamma_q = Z \gamma^p - 2 \im \eps^{lp} \gamma_l .
\end{equation}

The matrices $X_m$ and $Z$ are constructed from gamma matrices and objects with an even number of indices: the field strength~$\epsilon_{np}$, the Levi--Civita tensor, and the metric.  Thus, all terms in $X_m$ (resp.~$Z$) must involve an odd (resp.~even) number of gamma matrices.
A basis of $4\times 4$ matrices is given by $1$, $\gamma^m$, $\gamma^{mn}$, $\gamma^5\gamma^m$, $\gamma^5$.  Converting $\gamma^5$ to projectors $P_L$, $P_R$ we get the general form
\begin{equation}\label{X-expr}
  X_m = (A^-_{mn} P_L+A^+_{mn}P_R) \gamma^n , \qquad
  Z = \beta^- P_L + \beta^+ P_R + (1/2) C_{mn} \gamma^{mn} ,
\end{equation}
where $\beta^{\pm}$ are scalars and $A^{\pm}$, $C$ are two-index tensors built from~$\eps$ and the Levi--Civita tensor (and the metric), without any gamma matrices.  In addition $C_{mn}$ is antisymmetric.
We evaluate both sides of~\eqref{X-cond}, using $\gamma^q \gamma^n \gamma_r \gamma_q=4\delta_r^n$,
\begin{equation}\label{XZ-calc}
  \begin{aligned}
    X_n \gamma^q \gamma^n X^p \gamma_q
    & = (A^-_{nm}P_L+A^+_{nm}P_R) \gamma^m \gamma^q \gamma^n (A^{-pr}P_L+A^{+pr}P_R) \gamma_r \gamma_q
    \\
    & = (A^-_{nm}A^{+pr}P_L+A^+_{nm}A^{-pr}P_R) \gamma^m \gamma^q \gamma^n \gamma_r \gamma_q
    \\
    & = 4 A^{+pn} A^-_{nm} P_L \gamma^m + 4 A^{-pn} A^+_{nm} P_R \gamma^m ,
    \\
    Z \gamma^p - 2 \im \eps^{lp} \gamma_l
    & = (\beta^- \delta^p_m + 2 \im \eps^p{}_m + C^{+p}{}_m) P_L \gamma^m
      + (\beta^+ \delta^p_m + 2 \im \eps^p{}_m + C^{-p}{}_m) P_R \gamma^m .
  \end{aligned}
\end{equation}
where $C^{\pm}_{mn} \equiv C_{mn} \pm \im \tilde{C}_{mn}$ are \mbox{(anti-)}self-dual parts of~$C$.
The terms multiplying $P_L\gamma^m$ and $P_R\gamma^m$ in the two expressions must match:
\begin{equation}\label{ABBA-CDE-eqs}
  4 A^+ A^- = \beta^-\eta + 2\im\eps + C^+ , \qquad
  4 A^- A^+ = \beta^+\eta + 2\im\eps + C^- ,
\end{equation}
where we introduced the notation $FG$ for the matrix product $(FG)_{mn}=F_{mp}\eta^{pq}G_{qn}$.

To proceed further, we must understand better how the field strength can assemble into tensors and scalars $A^{\pm},\beta^{\pm},C^{\pm}$.  Firstly, the only invariant scalars of an antisymmetric two-tensor are $|\eps|^2=\eps_{mn}\eps^{mn}$ and the Pfaffian $\Pf\eps\equiv\frac{1}{8}\varepsilon^{pqrs} \eps_{pq} \eps_{rs}$, so $\beta^{\pm}$ are functions of these two invariants.
The tensors $A^{\pm},C^{\pm}$ are constructed from $\eps$ and the Levi--Civita tensor~$\varepsilon$, contracted using the metric.
If indices of two Levi--Civita tensors~$\varepsilon$ are contracted, they can be traded for the metric.
If indices of $\varepsilon$ and~$\eps$ are contracted they can be traded for the dual field strength thanks to $\varepsilon_{mnpq} \eps^{qr} = - \delta_m^r\ept_{np} - \delta_n^r\ept_{pm} - \delta_p^r\ept_{mn}$, and likewise for a contraction of $\varepsilon$ and~$\ept$.
Thus, $A^{\pm},C^{\pm}$ are constructed from the two-index tensors $\eps$ and~$\ept$, contracted via the metric.
A direct component calculation shows that $\ept = (\Pf\eps)^{-1}(- \eps^3 + (1/2) |\eps|^2 \eps)$, where $\eps^3=\eps\eps\eps$ is a matrix power.  This allows to recast $\ept$ in terms of powers of~$\eps$.
Furthermore, the Cayley--Hamilton theorem $\eps^4=(1/2)|\eps|^2\eps^2+(\Pf\eps)^2$ reduces all matrix powers of~$\eps$ to powers $\eps^k$, $k=0,1,2,3$.
Thus $A^{\pm},C^{\pm}$ are degree~$3$ polynomials in the matrix~$\eps$ with coefficients that depend on the scalars $|\eps|^2$, $\Pf\eps$.

Given the tensor structure of $A^{\pm}$, they commute as matrices, so that the two left-hand sides of~\eqref{ABBA-CDE-eqs} are equal.
In the resulting relation $(\beta^--\beta^+)\eta+C^+-C^-=0$ all terms have different symmetry properties, hence $C^+=C^-=0$ and $\beta^-=\beta^+$ (which we denote simply~$\beta$).  We are left with the matrix equation
\begin{equation}\label{AAB-eq}
  4 A^+ A^- = \beta \eta + 2\im\eps .
\end{equation}
For such a solution, inserting \eqref{X-expr} (and $Y=0$) into~\eqref{Mmn-X}, and using $\gamma^p\gamma_m\gamma^q\gamma_p=4\delta_m^q$ yields
\begin{equation}
  M_{mn} = -2 (A^+_{nm} P_R + A^-_{nm} P_L) .
\end{equation}
The evolution and constraint equations
\begin{equation}
  \Dslash\Psi_m = \im M_{mn} \Psi^n , \qquad
  \gamma^m \Psi_m = 0 , \qquad
  \Dcov^m \Psi_m = \frac{\im}{2} \gamma^m M_{mn} \Psi^n
\end{equation}
usefully decompose into equations on $\Psi^R_m\equiv P_R\Psi_m$ and $\Psi^L_m\equiv P_L\Psi_m$:
\begin{equation}
  \begin{aligned}
    \Dslash\Psi^R_m & = - 2 i A^-_{nm} \Psi^{Ln} , & \qquad
    \gamma^m\Psi^R_m & = 0 , & \qquad
    \Dcov^m\Psi^R_m & = -\im A^-_{nm} \gamma^m \Psi^{Ln} ,
    \\
    \Dslash\Psi^L_m & = - 2 i A^+_{nm} \Psi^{Rn} , & \qquad
    \gamma^m\Psi^L_m & = 0 , & \qquad
    \Dcov^m\Psi^L_m & = -\im A^+_{nm} \gamma^m \Psi^{Rn} .
  \end{aligned}
\end{equation}
A rescaling of one component (say $\Psi^L$) by a scalar function of $|\eps|^2/2$ and $\Pf\eps$ scales $A^{\pm}$ with inverse factors.
Overall, independent solutions of~\eqref{AAB-eq} depend $4$ scalar functions\footnote{One counts $1+4+4$ from $\beta$ and the coefficients of $1$, $\eps$, $\eps^2$, $\eps^3$ in $A^{\pm}$, minus $4$ from matching coefficients of $1$, $\eps$, $\eps^2$, $\eps^3$ in the equation, and $1$ from the rescaling invariance.} of $|\eps|^2/2$ and~$\Pf\eps$, leading to a large class of consistent evolution and constraint equations for spin-3/2 particles in a constant background field.

One should concentrate on solutions $A^{\pm}_{mn}, \beta$ of~\eqref{AAB-eq} that have no singularity at any value of the constant background field strength~$\eps$, as the system of constraints and evolution equations is otherwise ill-posed for some~$\eps$.
A convenient subset\footnote{Another interesting solution is $A^{\pm}=(1+\im\eps\pm \ept)/2$, and $\beta=1-|\eps|^2/2$, for which the left/right handed spinors are affected only by (anti-)self-dual parts of the field strength.  It is not of the form~\eqref{quite-general-Apm} because $A^+$ has a non-trivial kernel for some~$\eps$.} of solutions that depends on a maximal number of scalar functions $\beta,\alpha_1,\alpha_2,\alpha_3$ of $|\eps|^2/2$ and $\Pf\eps$ is to take $A^+$ to be invertible by expressing it as a matrix exponential (we have used the field redefinition to avoid a term $\alpha_0\eta$ in the exponential)
\begin{equation}\label{quite-general-Apm}
  \begin{aligned}
    A^+ & = \exp\bigl(\alpha_1 \eps + \alpha_2 \eps^2 + \alpha_3 \eps^3\bigr) = 1 + \alpha_1 \eps + (\alpha_2+\alpha_1^2/2)\eps^2 + O(\eps^3), \\
    A^- & = \frac{1}{4} \exp\bigl(-\alpha_1 \eps - \alpha_2 \eps^2 - \alpha_3 \eps^3\bigr) (\beta+2\im\eps) .
  \end{aligned}
\end{equation}
The choice that reproduces~\eqref{spin32 result} is the simplest one of this form, with $\alpha_1=\alpha_2=\alpha_3=0$: one gets $A^+ = - \eta/\sqrt{2}$ and $A^- = - (\eta + \im \eps)/\sqrt{2}$ (after rescaling by $-\sqrt{2}$), whose matrix product obeys $4A^+A^-=2\eta+2\im\eps$, which is~\eqref{AAB-eq} for a constant $\beta=2$.
This recovers and generalizes the equations in the previous section without going through a Lagrangian.

Finally, it is worth noting that the general form derived in this section always reproduces $g=2$ for the gyromagnetic ratio.
Applying $\Dslash$ to the equation of motion yields the second-order equation
\begin{equation}
    (D^2-\beta)\Psi_m +2Qi \epsilon_{mn}\Psi^n-\frac{1}{2}i Q \epsilon_{ab} \gamma^{ab}\Psi_m=0\label{secondorder}
\end{equation}
which does not depend on $A^{\pm}$, where for clarity we restored the charge $\epsilon\rightarrow Q\epsilon$. Without loss of generality, we choose vanishing electric field $\epsilon_{0i}=0$, while the magnetic field is defined by $\epsilon_{ij}=-\varepsilon_{ijk}B^k$ with $\varepsilon$ being the 3d Levi-Civita symbol. Note also that the covariant derivative is given by $D_m=\partial_m +i Q A_m$, and for a constant background, the potential is parameterized in terms of the spacetime coordinate $X$ as: $   A_n = -\frac{1}{2}(\epsilon X)_n$.  Going to momentum space, Eq.~\eqref{secondorder} reads
\begin{equation} \begin{aligned}
    \left(\mathcal{H}-Q A_0\right)^2 {\Psi}_{m}&=  \bigl[
 \bigl( \vec{p}- Q \vec{A}\bigr)^2+\beta\bigr] {\Psi}_{m}+Q \ep_{rs}\left(\mathcal{M}^{rs}\right)_{mn} {\Psi}^n\\&=\bigl[
 \bigl( \vec{p}- Q \vec{A}\bigr)^2+\beta\bigr] {\Psi}_{m}-Q \varepsilon_{ijk}B^k\left(\mathcal{M}^{ij}\right)_{mn} {\Psi}^n
\end{aligned}
\end{equation}
where $\mathcal{H} \equiv i\partial_0$, and $\left(\mathcal{M}^{rs}\right)_{mn}=\left(J^{rs}\right)_{mn}+S^{rs}\eta_{mn}$ are the generators of the Lorentz group,  with $\left(J^{rs}\right)_{mn}=-2i\eta_{m}{}^{[r}\eta^{s]}{}_n$, $S^{rs}=\frac{i}{2}\gamma^{rs}$. The generator of rotations for the spin-3/2 representation is $\left(S_k\right)_{mn}=\frac{1}{2}\ep_{ijk}\left(\mathcal{M}^{ij}\right)_{mn}$, so
\begin{equation} (\mathcal{H}-Q A_0)^2 {\Psi}_{m}=  \bigl[
 \bigl( \vec{p}- Q \vec{A}\bigr)^2+\beta\bigr] {\Psi}_{m}-2Q \Vec{B}\cdot\vec{S}_{mn} {\Psi}^n .
\end{equation}
The gyromagnetic ratio is recovered from the coefficient of the last term on the right hand side, which gives $g=2$, for any choice of solution of~\eqref{AAB-eq}.

\section{Massive charged spin-2 case}

A description of the propagation of charged massive spin-2 in 26 dimensions has been achieved by Argyres and Nappi~\cite{Argyres:1989cu,Argyres:1989qr} by deriving a Lagrangian for the first massive modes and subsequently by extracting  the equations of motion and constraint from the Virasoro algebra operator constraints~\cite{Porrati:2010hm}.  For the bosonic part of the open superstring, the Lagrangian in \cite{Benakli:2023aes,Benakli:2022edf,Benakli:2022ofz} features scalars, vectors and spin-2 physical degrees of freedom,
\begin{equation}
    \begin{aligned}
 \mathcal{L}={}&{\mathcal{\bar{M}}}_1\bigl( -2+\mathfrak{D}^2\bigr)\mathcal{M}_1+\bar{\mathcal{N}}_1\bigl( -2+\mathfrak{D}^2\bigr)\mathcal{N}_1\\&+\bar{\mathcal{C}}^m \mathfrak{D}^2\mathcal{C}_m+\mathfrak{D}^m \bar{\mathcal{C}}_m\mathfrak{D}^n {\mathcal{C}}_n-2 \mathcal{\bar{C}}^{m} \left(\eta_{mn}-\mathrm{i} \epsilon _{mn}\right) \mathcal{C} ^{n}\\&+2\bar{a}^ma_m-\mathrm{i}\epsilon_{mn}\bar{a}^m a^n+\mathfrak{D}^m\bar{a}_m\mathfrak{D}^n a_n+\frac{1}{\sqrt{2}}\Bigl[\bar{\tilde{F}}^{mn}(a)\left(F_{mn}(c)-\mathcal{H}_{[mn]}\right)+\text{h.c.}\Bigr]\\&-2\bar{c}^m c_m-\frac{2}{5}\mathfrak{D}^m \bar{c}_m\mathfrak{D}^n c_n+\Bigl[\bar{c}^m\Bigl(-\frac{2}{5}\mathfrak{D}_m\mathcal{H}+\mathfrak{D}^n\mathcal{H}_{nm}\Bigr)+\text{h.c.}\Bigr]\\&+ \frac{1}{2} \bar{\mathcal{H}}_{mn} \mathfrak{D} ^2h^{mn}+ \frac{1}{2}\mathfrak{D} ^{n} \bar{\mathcal{H}} _{mn} \mathfrak{D}_{k}h^{mk}-\bar{\mathcal{H}}^{(mn)}\mathcal{H}_{(mn)}+\mathrm{i}\epsilon^{nk}\bar{\mathcal{H}}_{mn}h_{k}{}^m+\frac{1}{10} \bar{\mathcal{H}}\mathcal{H} ,
\end{aligned}\label{compact-boson}
\end{equation}
where the rescaled spin-2 field is defined as
\begin{equation}
    \mathcal{H}_{mn}\equiv \left( \eta_{mk}-\mathrm{i}\epsilon_{mk}\right)h^{k}{}_n,\qquad \mathcal{H}=h .\label{rescaleH-def}
\end{equation}
We will focus on the equation of motion for spin-2. Let us define $\mathcal{P}_{m n}$ by  $\mathcal{P}_{m n}\equiv  \delta \mathcal{L}/\delta\bar{\mathcal{H}}_{mn}$, with
\begin{equation}
    \begin{aligned}
\mathcal{P}_{m n} =-\frac{1}{\sqrt{2}} \varepsilon_{m n k l} \mathfrak{D}^k a^l+\frac{2}{5} \eta_{m n} \mathfrak{D}^k c_k-\mathfrak{D}_m c_n & +\frac{1}{10} \eta_{m n} h+\frac{1}{2} \mathfrak{D}^2 h_{m n}-\frac{1}{2} \mathfrak{D}_n \mathfrak{D}^k h_{m k} \\
& -h_{m n}-\frac{3}{2} \mathrm{i} \epsilon_{k n} h^k{ }_m+\frac{1}{2} \mathrm{i} \epsilon_m{ }^k h_{k n} .
\end{aligned}
\end{equation}
Then the equation of motion of $h_{m n}$ is
\begin{equation}
2 \mathcal{R}_{m n}=(1-\mathrm{i} \epsilon)_m{ }^k \mathcal{P}_{k n}+(1-\mathrm{i} \epsilon)_n{ }^k \mathcal{P}_{k m}=0 .
\end{equation}
This relation is in general not invertible, because $\mathcal{R}_{m n}$ has 10 components by symmetry, while $\mathcal{P}_{m n}$ has 16 components. From this relation we cannot yet conclude $\mathcal{P}_{m n}=0$.
Meanwhile, the equations of motion of $a_m$ and $c_m$ are
\begin{equation}
\begin{aligned}
& \mathcal{E}_m \equiv 2 a_m-\mathrm{i} \epsilon_{m n} a^n-\mathfrak{D}_m \mathfrak{D}_n a^n-\frac{\mathrm{i}}{\sqrt{2}} \epsilon^{n k} \varepsilon_{m k p q} \mathfrak{D}^q h_n{ }^p+\sqrt{2} \mathrm{i} \ept_{mn} c^n=0 , \\
& \mathcal{F}_m \equiv-2 c_m+\frac{2}{5} \mathfrak{D}_m \mathfrak{D}_n c^n+\sqrt{2} \mathrm{i} \ept_{mn} a^n-\frac{2}{5} \mathfrak{D}_m h+\bigl(\eta^{n k}-\mathrm{i} \epsilon^{n k}\bigr) \mathfrak{D}_n h_{m k}=0 .
\end{aligned}
\end{equation}
We find the combination
\begin{equation}
-\frac{1}{\sqrt{2}} \varepsilon_{m n k l} \mathfrak{D}^k \mathcal{E}^l+\frac{1}{2}\left(\mathfrak{D}_m \mathcal{F}_n-\mathfrak{D}_n \mathcal{F}_m\right)=0=(1-\mathrm{i} \epsilon)_m{ }^k \mathcal{P}_{k n}-(1-\mathrm{i} \epsilon)_n{ }^k \mathcal{P}_{k m} .
\end{equation}
Adding the above equation to the relation (0.2) we find
\begin{equation}
-\frac{1}{\sqrt{2}} \varepsilon_{m n k l} \mathfrak{D}^k \mathcal{E}^l+\frac{1}{2}\left(\mathfrak{D}_m \mathcal{F}_n-\mathfrak{D}_n \mathcal{F}_m\right)+2 \mathcal{R}_{m n}=2(1-\mathrm{i} \epsilon)_m{ }^k \mathcal{P}_{k n}=0 .
\end{equation}
Multiplying by an inverse matrix $(1-\mathrm{i} \epsilon)^{-1}$ we can conclude\footnote{The rationale for using $\mathcal{P}_{mn}$ instead of $\mathcal{R}_{mn}$ has not been provided in \cite{Benakli:2023aes,Benakli:2022edf,Benakli:2022ofz}, hence the interest in presenting it in these proceedings.}
\begin{equation}
\mathcal{P}_{m n}=0 .
\end{equation}
By taking the trace and divergence of $\mathcal{P}_{mn}$ one can derive the Fierz-Pauli on-shell system for spin-2:
\begin{equation}
\begin{aligned}
\bigl( \mathfrak{D}^2-2\bigr)\mathfrak{h}_{mn}&=2\mathrm{i}\bigl( \epsilon_{km}\mathfrak{h}^k{}_n+\epsilon_{kn}\mathfrak{h}^k{}_m\bigr) , \\
 \mathfrak{D}^n\mathfrak{h}_{mn}&=0,\quad 
\mathfrak{h}=0 .
\end{aligned}\label{spin 2 eqs}
\end{equation}
To reach the above equations, one should introduce the following redefinition of the spin-2 field:
\begin{equation}
    \begin{aligned}
   \mathfrak{h}_{mn}\equiv{} & \frac{4}{3}h_{mn}-\frac{1}{3}\eta_{mn}h-\frac{\im}{2}\bigl(\ep_{m}{}^kh_{kn}+\ep_{nk}h^k{}_m\bigr)
                               +\frac{1}{3}\left(\mathfrak{D}_mc_n+\mathfrak{D}_nc_m\right)
      \\&-\frac{\im}{2}\Bigl(\ep_{mk}\mathfrak{D}^kc_n+\ep_{nk}\mathfrak{D}^kc_m-\ep_{mk}\mathfrak{D}_nc^k-\ep_{nk}\mathfrak{D}_mc^k+\eta_{mn}\ep^{kl}\mathfrak{D}_kc_l\Bigr)
      \\&-\frac{1}{4}\left(\ep_{mk}\ep^{lk}h_{nl}+\ep_{nk}\ep^{lk}h_{ml}+2\ep_{mk}\ep_{nl}h^{kl}-\eta_{mn}\ep^{kl}\ep^p{}_lh_{kp}\right)
      \\&+\frac{1}{96(2-\ep\ep)}\left[8\mathfrak{D}_m \mathfrak{D}_n h-6\ep_{mk}\ep^k{}_nh+12\im\ep_{mk}\mathfrak{D}^k\mathfrak{D}_nh-5\ep\ep\eta_{mn}h+\left(m\leftrightarrow n\right)\right]
      \\&-\frac{1}{8\sqrt{2}(2+\ep\ep)}\Bigl[-4\im\bigl(\ept_{mk}\mathfrak{D}^k\mathfrak{D}_n+\ept_{nk}\mathfrak{D}^k\mathfrak{D}_m\bigr)\mathfrak{D}_la^l+5\left(\ep\ept\right)\eta_{mn}\mathfrak{D}^ka_k
      \\&\qquad -2\left(\ep\ept\right)\left(\mathfrak{D}_m\mathfrak{D}_n+\mathfrak{D}_n\mathfrak{D}_m\right)\mathfrak{D}_ka^k+8\bigl(\ept_{mk}\ep_{ln}\mathfrak{D}^k\mathfrak{D}^l +\ept_{nk}\ep_{lm}\mathfrak{D}^k\mathfrak{D}^l \bigr)\mathfrak{D}^pa_p\Bigr] .
    \end{aligned}\label{sp2ABdef-comp}
\end{equation}
Notably the spin-2 equation of motion~\eqref{spin 2 eqs} can be derived by squaring the Dirac equation for the spin-3/2 field.

\section{Summary}

The equations of motion and constraints of the bosonic fields of integer spin $s$ read
\begin{equation}
\begin{aligned}
\bigl( \mathfrak{D}^2-M^2\bigr)\mathfrak{h}_{m{ n_2\cdots n_s}} & = 2\im\epsilon_{k(m}\mathfrak{h}^k{}_{ n_2\cdots n_s)} ,
\\ \mathfrak{D}^m\mathfrak{h}_{m n_2\cdots n_s} & = 0 ,
\\ \mathfrak{h}^m{}_{m n_3\cdots n_s} & = 0 ,
\end{aligned}
\end{equation}
which are a generalization  of the results of \cite{Argyres:1989cu} and are also shown in \cite{Porrati:2010hm}. Here, the spin $s$ field is represented by a symmetric traceless tensor $\mathfrak{h}_{n_1 n_2\cdots n_s}$. For the fermion of spin $s+1/2$, represented by a symmetric tensor-spinor $\Psi_{n_1n_2\cdots n_s}$, we get
\begin{equation}
\begin{aligned}
\left(i
\slashed{\mathfrak{D}}+M\right) \Psi_{mn_2\cdots n_s} & = - \im  \frac{2}{M} \epsilon^{}_{k (m} {\Psi}_{n_2\cdots n_s)}^{Lk} ,
\\ \mathfrak{D}^m\Psi_{mn_2\cdots n_s} & = \frac{1}{2M}\gamma^m\bigl(\epsilon^{}_{k(m}+\im\Tilde{\epsilon}^{}_{k(m}\bigr)\Psi^{k}{}^{}_{n_2\cdots n_s)} ,
\\ \gamma^m \Psi_{m n_2\cdots n_s} & = 0 ,
\end{aligned}
\end{equation}
where the $L$ superscript denotes a projection on left-handed spinors by $P_L=(1+\im \gamma^5)/2$, while a right-handed projection is also possible by $P_R=(1-\im\gamma^5)/2$.

We have shown that the equation of motion and constraints for the massive charged spin-3/2 can be recovered without appealing to the Lagrangian. The present available Lagrangian suffer from the presence of extra fields with lower spins both for the spin-3/2 and spin-2 fields. The challenge remains to exhibit alternative versions without these fields.

\section*{Acknowledgements}

K.B. gratefully acknowledges Nathan Berkovits and Matheus Lize. K.B. and W.K. also extend their appreciation to Cassiano Daniel.

\end{document}